\documentclass[10pt,conference]{IEEEtran}
\IEEEoverridecommandlockouts

\usepackage{listings}
\usepackage{float}
\usepackage{verbatim}
\usepackage{stfloats}
\usepackage{textcomp}
\usepackage{booktabs, tabularx}
\usepackage{rotating}
\usepackage{xcolor}
\usepackage{cite}

\usepackage{fancybox}
 \usepackage{multirow}

\usepackage{hyperref}

\usepackage{url}
\usepackage{hyperref}

\usepackage[caption=false,font=footnotesize]{subfig}
\usepackage{balance}
\usepackage{footnote}
\usepackage{xspace}
\usepackage{array}
\usepackage{comment}
\usepackage{adjustbox}
\usepackage{threeparttable}

\newcommand{\phead}[1]{\vspace{1mm} \noindent {\bf #1}}

\usepackage{tcolorbox} 
\usepackage{textcomp} 

\definecolor{hzcolor}{RGB}{10, 186, 181}

\newcolumntype{M}[1]{>{\centering\arraybackslash}m{#1}}

\definecolor{dkgreen}{rgb}{0,0.6,0}
\definecolor{gray}{rgb}{0.5,0.5,0.5}
\definecolor{mauve}{rgb}{0.58,0,0.82}

\usepackage{array}
\newcolumntype{$}{>{\global\let\currentrowstyle\relax}}
\newcolumntype{^}{>{\currentrowstyle}}

\definecolor{pgrey}{rgb}{0.46,0.45,0.48}

\newcolumntype{C}[1]{>{\centering}m{#1}}

\def\BibTeX{{\rm B\kern-.05em{\sc i\kern-.025em b}\kern-.08em
    T\kern-.1667em\lower.7ex\hbox{E}\kern-.125emX}}
\begin{document}

\title{Understanding the Challenges and Assisting Developers with Developing Spark Applications}

\author{\IEEEauthorblockN{Zehao Wang}
\IEEEauthorblockA{\textit{Department of Computer Science and Software Engineering} \\
\textit{Concordia University}\\
Montreal, Canada \\
w\_zeha@encs.concordia.ca}}



\maketitle

\begin{abstract}

To process data more efficiently, big data frameworks provide data abstractions to developers. However, due to the abstraction, there may be many challenges for developers to understand and debug the data processing code. To uncover the challenges in using big data frameworks, we first conduct an empirical study on 1,000 Apache Spark-related questions on Stack Overflow. We find that most of the challenges are related to data transformation and API usage. To solve these challenges, we design an approach, which assists developers with understanding and debugging data processing in Spark. Our approach leverages statistical sampling to minimize performance overhead, and provides intermediate information and hint messages for each data processing step of a chained method pipeline. The preliminary evaluation of our approach shows that it has low performance overhead and we receive good feedback from developers. 

\end{abstract}


\section{Introduction}\label{sec:introduction}

Big data processing frameworks, such as Apache Spark~\cite{spark}, provide abstractions to developers so that they can process data more efficiently and easily. However, there may be many problems in the Spark development process. Prior research has focused on what big data topics are popular and what topics are more difficult~\cite{Bagherzadeh:2019:GBL:3338906.3338939}. However, there is a lack of understanding of the challenges that developers encounter during Apache Spark development and what kind of supports are needed.

In this paper, we follow a three-phase sequential exploratory strategy~\cite{reactiveDebug, creswell2017research, Hanson2005MixedMR} to study the challenges that developers encounter when debugging and developing Spark applications.
First, we conduct a qualitative study on 1,000 Spark-related questions on Stack Overflow, which reaches a 95\% confidence level and 3\% confidence interval. We study the questions, answers, comments, and the associated  code snippets. 
The results show that problems related to Data Processing and Spark API Usage are the most common challenges that developers encounter -- accounting for 63\% of the studied posts.
Second, based on our findings, we design an approach to help developers debug and understand big data application execution. We implement our approach as an independent package for PySpark, which is the official Python version of Spark. Our approach leverages statistically sampling to minimize performance overhead, and provide intermediate processing information and hint messages for each data processing step of a chained method pipeline. 
Finally, we conduct a preliminary evaluation of our approach on six benchmarking programs and the results show that the performance overhead is low. In the future, we plan to conduct a user study to evaluate the effectiveness of our approach.

\phead{Paper organization.} Section~\ref{sec:motivation} presents the challenges in developing Spark applications that we uncover from Stack Overflow posts. Section~\ref{sec:methodology} presents the design of our approach and preliminary results. Section~\ref{sec:future} discusses the future work. Section~\ref{sec:conclusion} concludes the paper.

\section{Challenges in Developing Spark Applications}\label{sec:motivation}
\phead{Approach.} We collected 12,217 Spark-related questions from the 2014 year to the 2019 year that have an accepted answer and code snippet from Stack Overflow. We conduct a qualitative study on a statistically significant sample of questions and their associated answers. More specifically, we randomly sample 1,000 questions, achieving a confidence level of 95\% and a confidence interval of 3\%. We performed a lightweight open coding-like process to find categories of challenges that developers encounter by manually labeling 1,000 posts.

\phead{Results.}
We derive six categories of challenges that developers encounter: Data Processing, Spark API Usage, Configuration, Data Sources, Performance and Logging, and Other. We find that the most common challenge that developers encounter is related to Data Processing (43.2\%). Data processing is the most common functionality of Spark. However, when the code runs into an unexpected result, developers may have trouble identifying which data processing method causes the issue as developers cannot see the intermediate data processing result step by step. For example, a developer on Stack Overflow had a data transformation application~\cite{SO4}. The developer wishes to display step-by-step execution results to test his application, because every time the program runs for half an hour, an exception will be thrown. The suggested answer is to sample the data and test the application locally instead of on the cluster.

The second most common challenge is related to Spark API Usage (19.1\%). Since Spark integrates the functional programming paradigm in its API design to abstract big data processing, sometimes developers may not be familiar with the working mechanism of an API and can use the API incorrectly. In addition, some data processing methods contain optional parameters that provide different ways to process data, but developers may not be aware of such options. For example, a developer asked a question on Stack Overflow that the Spark API did not get the expected result~\cite{SO5}. The developer wished to convert the SQL string to the Spark DataFrame API format, but when he uses the DataFrame API, the result is different. By using SQL, Spark sorts the two columns according to \texttt{order by}. Although the descending order parameter is specified, according to the working principle of SQL, this parameter is only valid for the second variable, and the first variable is arranged in ascending order according to the default parameter value. This developer made an error while using Spark's DataFrame API. The developer assigned \texttt{False} to the \texttt{ascending} parameter. In this case, Spark will sort the data in descending order of the two columns. Developers can avoid this kind of problem by using the API and parameters correctly. In this case, providing some hints on anomalous data processing results and parameter usage may help developers understand and debug data processing in Spark. Overall, questions related to Data Processing and Spark API Usage are the most common challenges that developers encounter -- accounting for 63\% of the studied questions.

 We also find that developers often encounter challenges in configuring Spark (15.1\%) and its interaction with other data sources (11.4\%). Due to the high configuration flexibility and complexity of Spark, developers often encounter difficulties in the Spark configuration process. In addition, Spark provides APIs to read or store data from various data sources, such as databases. Developers may encounter problems in this process. We find that 5.5\% of the questions are related to performance and logging issues in Spark deployment. Developers may encounter problems in how to improve the performance of Spark applications and properly configure or use logs. Finally, there are some questions (5.4\%) that we categorize into the other category, which includes known and unresolved bugs in Spark or questions that are related to programming language syntax.

Based on our manual analysis of Stack Overflow questions, we further distill three debugging and monitoring challenges that developers have when using Spark: 
\begin{itemize}
    \item {\bf Challenge 1:} Data processing in Spark usually involves a series of steps to transform the raw data into an understandable/usable format. However, it is usually impossible for developers to know the intermediate results (or state) of the processed data, which raises challenges in debugging data processing steps.

    \item {\bf Challenge 2:} Developers may not be familiar with the input requirement or parameter values of the data processing methods. 

    \item {\bf Challenge 3:} There is a lack of tooling supports that show the data processing details with low runtime overhead to monitor and help debug data processing steps.
\end{itemize}

\section{Design and Preliminary Results}\label{sec:methodology}
To address the challenges that we uncovered in Section~\ref{sec:motivation}, we present the design of our approach, which assists developers with understanding the data processing execution. Our approach provides the intermediate information (e.g., data changes and states, and anomalies in the data processing) from each of the executed Spark methods.


\textbf{Recording intermediate information:} Our approach records information of the data after each data processing method is executed during runtime (e.g., data.filter('age \textgreater 18').join(data2, 'age')). It records two types of information: {\itshape data state} and {\itshape data processing}. For {\em data state}, it records the information of the data state before and after each method. For {\em data processing}, Our approach records a small sample of the data before and after applying the data processing method for showcasing and debugging.


\textbf{Hints for application:} As we found in Section~\ref{sec:motivation}, developers sometimes may not be familiar with the parameters used in data processing methods. Moreover, bugs that developers face do not always run into exceptions or failures, but may also be related to incorrect calculation or data. For different functions, Our approach will provide two types of hints: {\itshape method parameters} and {\itshape anomalous data processing results}. More specifically, our approach will check the values of the parameters given to the data processing method and analyze abnormal situations to give developers hints.


\phead{Preliminary Results.} We implement six Spark benchmarking programs based on a prior study~\cite{10.1145/2884781.2884813} for our experiment. The preliminary results show the overhead of our approach is small (i.e., within seconds for 5GB of data), and the overhead remains stable even when the data size is increased by 100 folds (i.e., from 50MB to 5GB). Our initial finding shows that our approach has a small performance overhead and good scalability. 


\section{Future Work}\label{sec:future}
In the future, we plan to design a user study to evaluate the effectiveness of our approach. We plan to design several debugging tasks based on the real Stack Overflow questions that we studied in Section~\ref{sec:motivation}. 
Each participant will be assigned with all the tasks and is required to debug three tasks with the help of our approach and debug another three tasks without help. We will record the time it takes for each participant to finish each task, and ask the participant to rank the usefulness of our approach. We receive good preliminary feedback from our initial user study. 

\section{Conclusion}\label{sec:conclusion}
In short, this paper makes the following contributions: 1) We conduct an empirical study of Spark-related questions on Stack Overflow and classify the major challenges
that Spark developers encounter: data processing and API usage. 2) We propose an approach to assist developers with understanding and debugging data processing errors in Spark. 3) The preliminary result shows that our approach has a small runtime overhead based on six Spark benchmarking programs. In the future, we plan to conduct a user study to evaluate the effectiveness of our approach.
\balance

\bibliographystyle{./bibliography/IEEEtran}
\bibliography{main}

\end{document}